# Understanding the effect of loss-framing in insurance purchase decisions – using a *game of life* like interface

Kunal Rajesh Lahoti

Kunal.lahoti@research.iiit.ac.in

The International Institute of Information Technology - Hyderabad

Professor CR Rao Rd, Gachibowli, Hyderabad, Telangana 500032

Shivani Hanji

Shivani.hanji@alumni.iiit.ac.in

The International Institute of Information Technology - Hyderabad

Professor CR Rao Rd, Gachibowli, Hyderabad, Telangana 500032

Pratik Kamble

Pratik.kamble@alumni.iiit.ac.in

The International Institute of Information Technology - Hyderabad

Professor CR Rao Rd, Gachibowli, Hyderabad, Telangana 500032

Kavita Vemuri

kvemuri@iiit.ac.in

The International Institute of Information Technology – Hyderabad

Professor CR Rao Rd, Gachibowli, Hyderabad, Telangana 500032

Data Availability Statement

The data and application developed will be provided on request to researchers.

Funding statement

This was not a funded project. Resources of the lab were utilized.

Conflict of Interest Statement

The authors declare that the research was conducted in the absence of any commercial or financial relationships that could be construed as a potential conflict of interest.

**Abstract**

This study investigates the impact of loss-framing and individual risk attitude on the willingness- to purchase insurance products, utilizing a game-like interface in a choice architecture. The application presents events as experienced in real-life , as a young adult searching for employment to retirement. Both financial and emotional loss-framing events are followed by choices to purchase insurance. The participant cohorts considered were undergraduate students and older participants, the latter group was further subdivided by income and education. From the within-subject analysis, the preliminary findings reveal that the loss framing effect on insurance consumption is higher in the younger population though contingent on the insurance product type. Health and accident insurance shows a negative correlation with risk attitudes for the younger age and a positive correlation with accident insurance for older participants. Risk attitude and life insurance products willingness-to-purchase showed no dependency. The findings elucidate the role of age, income, family responsibilities, and risk attitude in purchasing insurance products. Importantly, it confirms the correlation between age and the effect heuristics of framing/nudging.

# 1 | INTRODUCTION

Insurance products are positioned as a safety net to protect an individual, family ([1]), or property from financial shocks. Depending on the type of insurance, it supports the future financial demands of family members in the event of the insured's death/disability/retirement (life insurance), medical expenditures (medical insurance), property loss due to natural calamities or theft and accident-related losses (accident insurance). By design, insurance is a diversified financial product to protect individuals from financial losses or as an investment plan in a tax-sheltered environment ([2]). Given the long-term gains from insurance, there is an expectation that these products would have high consumption rates ([3]), particularly on utility maximising by considering all constraints ([4]). That is, one is expected to weigh the risks, the probability of an unfortunate event and the subsequent loss and make a rational choice. This understanding was supported by earlier theoretical models that have posited that risk-averse rational agents at certain personal wealth levels purchase more insurance [5]. However, as demonstrated in experiments conducted by behavioural economists ([6]; [7]; [8]; [9]), cognitive bias based on under-estimating the probability of risks ([10]; [11]; [12]), search costs in information gathering ( [13]) and low self-experiences of facing an unfortunate event ( [14]) lead to decisions that classical rational choice theories cannot fully explain. Other influencing factors include the role of individual cognitive processing load, the socio-economic-cultural determinants and risk-taking propensity. Risk-aversion was inferenced with the underlying premise that highly risk-averse individuals are more likely to purchase insurance, as they seek to protect themselves against potential losses ([15]). Our study takes into cognizance the review by Harrison (2019) [16] positions the theories on risk aversion, time preference, subjective beliefs, information processing while also highlights the challenges of interpretation of behavioral dynamics towards insurance, and in the long run the effect on policy interventions.

The Indian insurance consumer size is small though growing, with penetration at 4.2% (FY2021). Life insurance is the most popular product at 3.2%, while non-life insurance consumption is 1% ( [17]). Demand is also a function of one' wealth, prospects, and alternate choices with better returns (for example: real estate, gold). Though insurance is a long-term gains product, the Indian population with strong savings habits ([18]) invests mainly in precious metals like gold or real estate. Other reasons for low penetration cited are awareness ([19]), individual risk-taking level, sustained trust in institutions, financial literacy ([20]) and access to the financial market. The findings ([21]) of long-term insurance schemes perceived as risky due to the uncertainty associated with insurance payouts, lapsing, and premium increases, could also be applicable to the Indian consumer. A significant factor is the low-key marketing by the largely government-held insurance company, especially for medical insurance. There is an uptake due to government-provided subsidies for health and compliance requirements for vehicle insurance, in the recent years. Standard marketing strategy by companies and governments is by contextual framing, nudges, and rewards to influence choices ([22]) or explicitly position the risks for better situational awareness. And methods to communicate the product details and information on risk management are by visual aids with explicitly displayed probabilities of an event occurrence. The market factors specifically influencing insurance purchase are inferred from traditional surveys or publicly available data released by insurance companies. The analysis focuses on whole-life or term-life insurance choices to examine within-subject differential choices for each type of insurance to study the temporal changes in demand ([23];[24];[25]). Survey data collection has various drawbacks, for example, asking direct questions leads to socially accepted responses, errors due to language comprehension, lack of standard format, and notably, for many, a survey is perceived as tedious. Second, findings on insurance consumption in western societies ([26];[27]) might not be directly applied in other countries. Hence, it is imperative to test situated contexts, individual risk attitudes and the role of socio-economic conditions. Few studies have combined an individual's response to different insurance products, as a function of risk attitude and loss-framing. Experimental laboratory studies have also not explicitly presented real-life events as situational placement.

Towards addressing the above listed gaps, the methodological approach for this study is: a) a *game of life* like application that presents a real-world simulation of life events. In selecting a game-like application, we considered the findings from Cai & Song [28], who investigated whether insurance games could increase the use of weather insurance and the work of Patt et al [29], who similarly used field games, b) presenting information on the current financial situation (income, expenditure, savings etc.) and a clear insurance contract, we hope to reduce cognitive load and other confounding factors in decision-making while also attempting to address the discussions on bounded rationality, c) presenting a variety of insurance products (life, health, accident) and d) loss framing ([30]) to make participants aware of the advantages of insurance policies to analyze its effect on customers to acquire insurance [31]. Loss framing is a cognitive bias that influences how people perceive and respond to decisions or events in terms of prospective losses or profits. It is important in altering people's decision-making processes when it comes to insurance consumption ([32]). The effectiveness of loss framing is recognized by insurance firms, who include it into their marketing and communication campaigns. They emphasize anecdotes or data regarding people who incurred huge financial problems because of unforeseen catastrophes yet did not have insurance to protect them. The considered methodology has the benefit of informing participants about the significance of insurance through contextual framing rather than explicit instructions.

The social-economic factor's influence were by administering to three groups: a) a student cohort with no immediate financial responsibilities, b) an older age group with financially dependent family members and c) young professionals in high-paying jobs and higher financial prospects. The participant groups were selected to consider age, socio-economic prospects, and family responsibilities. The analysis aimed to understand factors loading risk perception and the role of financial stability. And the individual's risk attitude was measured by the DOSPERT questionnaire ([33]) and Willingness to Purchase (WTP) a specific insurance. The DOSPERT questionnaire is commonly administered to correlate individual differences in risk-taking behavior to real-world choices.

The analysis includes the following:
1. The choice to purchase a particular insurance product and the correlation to age & financial responsibilities.
2. Correlations of individual risk attitudes/traits and insurance consumption.

The interpretation is based on the fundamentals of behavioural economics: a) loss incurrence or negatively framed information, rooted in the prospect theory, b) regulatory focus theory – wherein the goals and motivation can change as the feedback is provided and c) choice in the type of insurance (life, health, accident).

We found that the loss framing effect exists for individual health, family health and accident insurance products across groups, except for a reverse trend for individual health insurance in the younger population, consistent with Sebastian's [34] work. The difference in family health insurance consumption before and after 'loss' in family health insurance is larger for younger participants than for older participants, while the reverse is true for accident insurance. From the correlation analysis, we observed that risk attitude, family health and accident insurance show a significant (negative correlation coefficient) relationship in younger population, while a significant positive correlation was evident for accident insurance in older participants .

## 2 | Background

Utility theory as a framework is valuable in analyzing participants' insurance choices in uncertainty. Demand for insurance is one of the significant outcomes of utility maximization research ([35]; [36]; [37]), and an integral part is risk preferences, also known as expected utility theory (EUT). The theory considers individuals' probability measures to be particularly important when deciding on an insurance purchase ([38]). We purchase full insurance if and only if premiums are actuarially fair, that is, equal to the expected payout from an insurance company ([39]). Under uncertainty, EUT plays a significant role in determining insurance premiums. The rational choice basis of EUT falls short of explaining why people do not buy insurance for high-intensity losses that occur with low probability and buy insurance for events with moderate loss and high probability ([40]; [41]; [42]). Prospect theory developed to extend EUT, introduced two significant innovations: reference dependence and probability weighting, and when applied to insurance, it implies that people buy insurance for reasons other than the certainty it provides ([43];[44]).

Investigation into the factors affecting decision-making by individuals has studied the role of cognitive processing using behavioural paradigms like "choice architecture" ([45]). This proposes that poor decision architecture could encourage users to favor short-term benefits over substantial long-term gain, resulting in irrational choices. In a few cases, the inverse is also noticed, wherein people tend to overestimate the likelihood of hazards and pay a higher premium in insurance purchases due to the complex composition of the decision architecture ([46]). For example, it was found that people assigned a low probability to high-intensity damages (caused by natural disasters significantly), and the benefit-cost ratio was underestimated ([47];[48]).

The role of one's negative experiences in insurance purchases and claims cannot also be discounted. However, an important factor is low financial literacy restricting one's ability to project future income, estimate expenses and design savings plans accordingly ([49];[50];[51];[52]). Another critical concept is consumer smoothing ([53]), which says that risk-averse households will employ any means possible to avoid a significant drop in consumption (such as taking children out of school), even in phases of low inflow of income. When applied to insurance behaviour, Bailey [54], found that the welfare gain from social insurance is determined by the product of the percentage consumption drop caused by the shock and the coefficient of relative risk aversion. There is also a strong hypothesis that an individual's risk-averse attitudes might result in a purchase of more insurance products

([55];[56]). while life events like new parenthood result in a demand for life insurance ([57]). Studies ([58];[59];[60];[61]) have examined life insurance purchases across different countries and found socio-economic status, family size, and demographics as significant determinants. To assign weightage to each of the listed factors, it is essential to consider the differences as a function of the type of insurance. Though there was no significant association between demand for long-term care insurance ([62]) or life insurance ([63]) consumption, the two products are functionally different. Life insurance ensures a higher return at the end of the term or to the next of kin in case of premature death, while medical and accident insurance cover the expenses. As a rational choice, life insurance builds up a cash value over time and thus functions as a "savings and insurance" tool. Moreover, because it "guarantees" payment of a death benefit, it may appear to be "safer" ([64]). As a result, life insurance can be viewed as a risk-free investment option, while other products are expenditures and risk-prone. In this study, we attempt to analyze individual choices with a unique experimental paradigm when options of economic security of insurance as a long-term investment and the financial loss due to mishap are offered through loss framing ([65]). That is, presenting relevant information or setting the context for situational awareness to support decision choices.

In India, insurance penetration of products such as accident and health insurance remains low ([66]; [67]). The primary reason for poor penetration is underestimation of events with a low probability of occurring ([68]). Several behavioral economic theories and empirical studies have been proposed to explain why people misjudge these risks and fail to plan for them ([69]), which need to be examined for validity on Asian Indian populations. The absence of loss experience has also been linked to low insurance demand, and influenced by low risk estimates ([70]).

**3 |METHODOLOGY**

3.1 Participants

Data was collected from three sets of participants:

1. Group A: 97 undergraduate students (age: 18 and 25; mean = 20.7±1.6). The participants were financially dependent on their parents.
2. Group B:- 42 older adults (age: 35 and 55, mean=41.3 ±4.3 ), employed as security guards and administrative staff at the institute with a salary range of INR 20,000 – 30,000 per month.
3. Group C: - 65 participants similar in age range to group B (mean: 38.4± 2.9), but for the education levels (IT professionals mainly) and a salary in the range of INR 50,000 to 70,000 per month.

The gender of the participants was not considered as a factor in the analysis and hence not recorded.

3.2 Experimental Design

A web-based game-like application is designed and developed using *Vue (an open-source front-end application to build interfaces using Javascript)*. The interface (Figure 2) presents an individual's typical life stages (flow chart in Figure 1) condensed into 12 months, starting with a job as a young adult, getting married, having children, old age and retirement. Events requiring payment to a family member, self-expenses and insurance options were presented at regular intervals (Figure 2a-d). The game moves linearly, with the player having the control to move to the next screen post making a choice from the options presented. No timer was included to remove effects from time anxiety.. The navigation is by a right scroll design for the progression of events. The options are forced-choice binary (yes/no) decisions or fixed amounts to pay (for a family event for example) or receive. To emulate real-life conditions, an amount as salary is credited into the player's account at each turn (a turn represents a month). Depending on the frequency and type of event, the balance (in savings, deposits, etc.) is updated and presented for quick reference (Figure 2e). The monthly savings are displayed after an automatic deduction of premium amounts (for any insurance purchased) and a fixed monthly living expense. As per design, the player is presented with different events, in addition to insurance products, and must make a choice from the 2-options provided. To emulate a loss-framing, a medical emergency to self or a family member and an accident is randomly generated. The accident and health insurance options are presented twice, pre-post and mishap, and subsequent hospital expenses are deducted. The instances in which people incur a monetary loss without insurance coverage act as negative feedback. The information is presented in text or audio format in Hindi, English, and Telugu to reduce confounds due to language comprehension. The paradigm is designed to generate awareness and influence the absorption of new financial products that require presenting uncertainty and understanding of the benefit-to-loss ratio. The secondary objective of the game is for awareness creation on financial management - that is, manage debt and current balance, and encourage savings.

We employ a Within-subject design in our experiment, to help us investigate the effect of loss-framing on insurance consumption (willingness to purchase or WTP). Second, to make inferences on the differential choices to insurance products, a within-subject design was deemed optimal. We do not provide within-game incentives, after due analysis on the following lines: a) Within-game incentives undermine the reliability and validity of the research results, b) eliminate the effect of moral dilemma ([71]) and c) incentives may attract individuals primarily motivated by financial gains rather than genuine participation in the experiment. The limitations of not offering incentives could be participants engagement, and to offset this a small participant fee was provided.

3.3 Experiment paradigm

The participants signed a consent form, and an honorarium of INR 50 was paid. The experiment was conducted in-lab and online (groups C & A), with identical procedures and content. The application was shown on a desktop screen for the in-lab experiment while online participants logged into the application. As part of the instructions, the game's interface, the rules of the game and the information provided as feedback at each step was explained.  A total of 25 events (Figure 1) were presented in the game. Post the gameplay, all participants completed a survey with risk-trait questions ([72]) followed by personal details of age, any present significant health condition (yes/no choice only) and financial commitments in real-life like loans and family dependents.

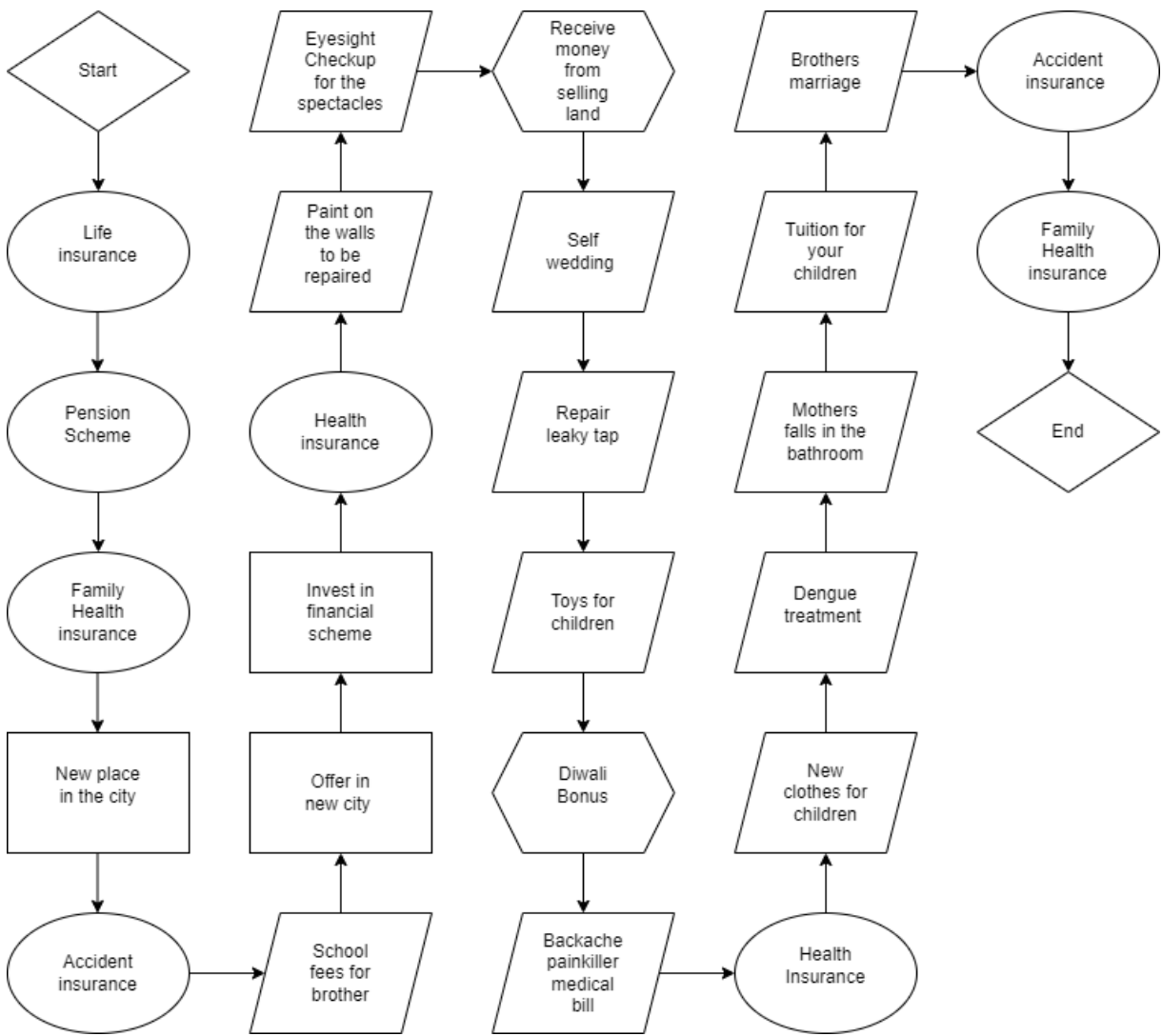

Figure 1: The flow chart displays all the events in the game. In the application, there are three basic types of events: -

A. Events that cause a drop in net worth or as losses (parallelogram).
B. Events that result in a rise in net worth (hexagon).
C. Insurance-related events (oval).
D. The change from the status quo (rectrangle).

Figure 2 presents sample screenshots of the application. The toolbar on the top has a menu button for 'language' option and the name of the project (in Hindi) translated from 'game of life'. The updated financial information is presented on the right panel (figure 2c right panel). The option to listen to the 'choices' or 'information' is provided by the audio button. All insurance products details and contracts as rolled out by government agencies (figure 2a) are presented as a series of choices interspersed with real-life events requiring expenditure (figure 2b) or cash inflow (figure 2c).

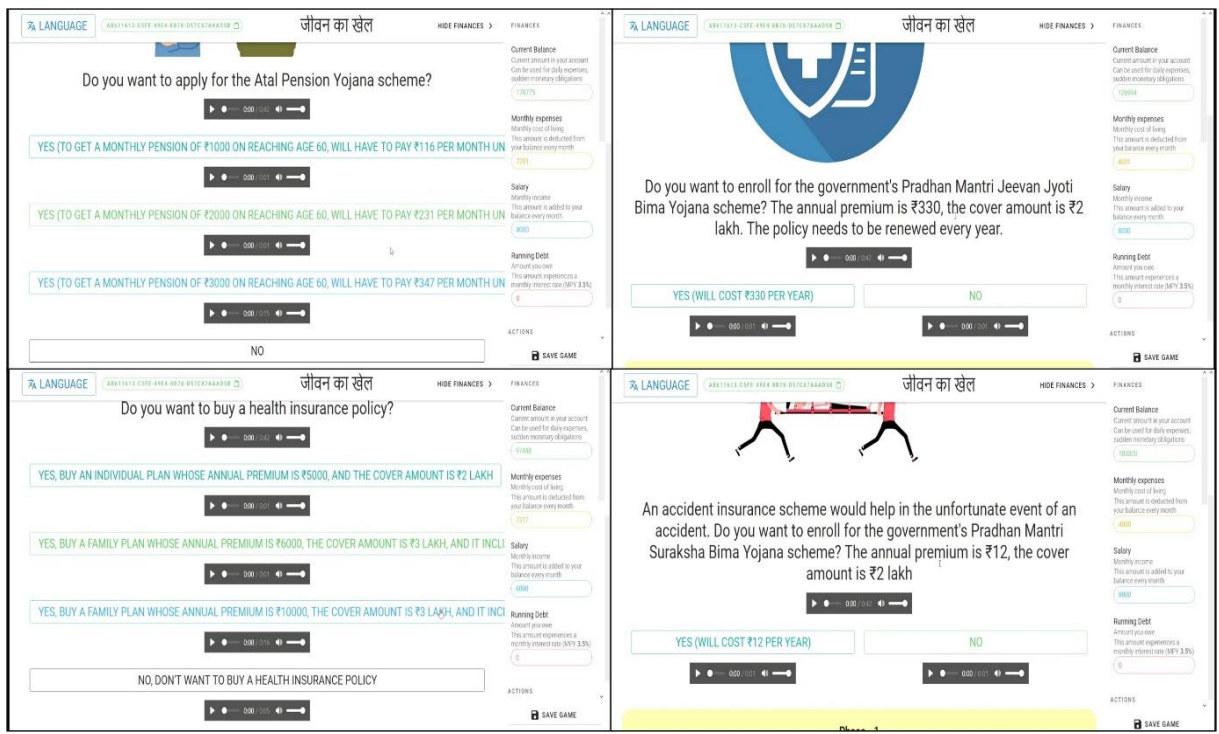

Figure 2a: Figure displays the various insurance coverage available to participants in the game. The illustrations illustrate a pension plan (top left), a life insurance policy(top right), a health insurance policy (bottom left), and an accident insurance policy (bottom right). The Atal Pension Yojana [73]is a government-backed pension system in India that aims to policy holders with a sustainable income during their retirement years. The Pradhan Mantri Jeevan Jyoti Bima Yojana (PMJJBY) [74] is an Indian government-sponsored life insurance plan. The Pradhan Mantri Suraksha Bima Yojana (PMSBY) [75] is India's government-sponsored accident insurance plan.

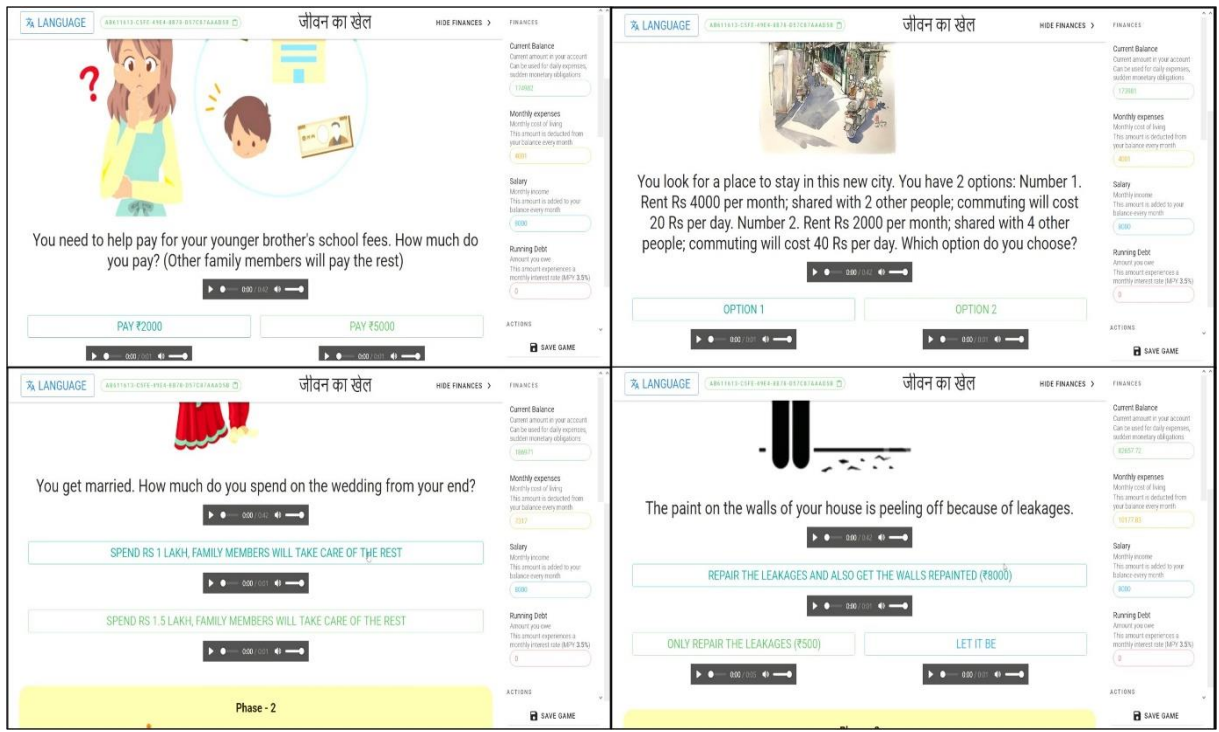

Figure 2b: The figure shows different events leading to expenses. An instance when the player has to decide on the amount to spend on brother's education fees (top left), making a choice between two house rental options (top right), budgeting for self-wedding (bottom left), and cost for house maintenance (bottom right).

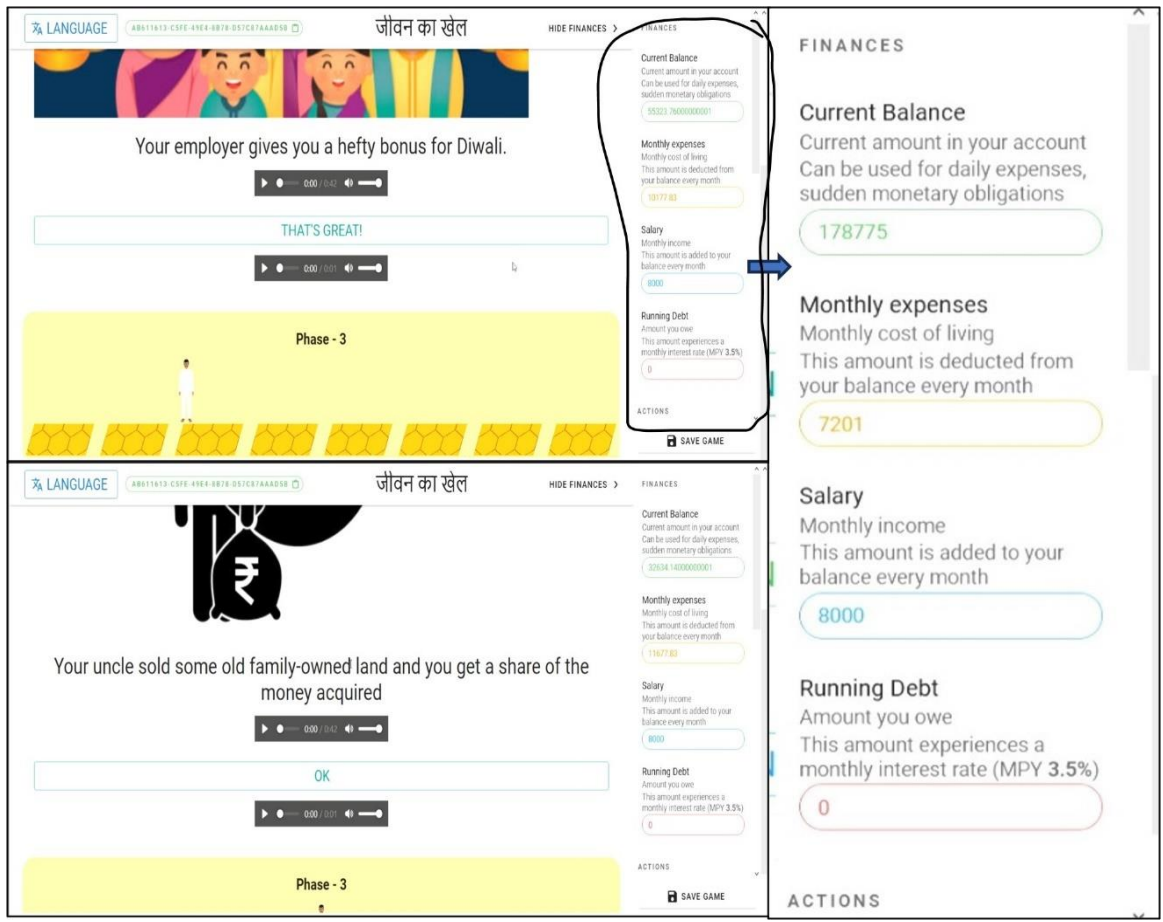

Figure 2c: The figure depicts several cash inflow events in the game. The image on the top left shows the Diwali bonus, and the image on the bottom left shows the capital gain from the sale of family land. The graphic on the right depicts the financial information made available to participants through the game. Finance information includes current balance, monthly balance, salary, and running debt.

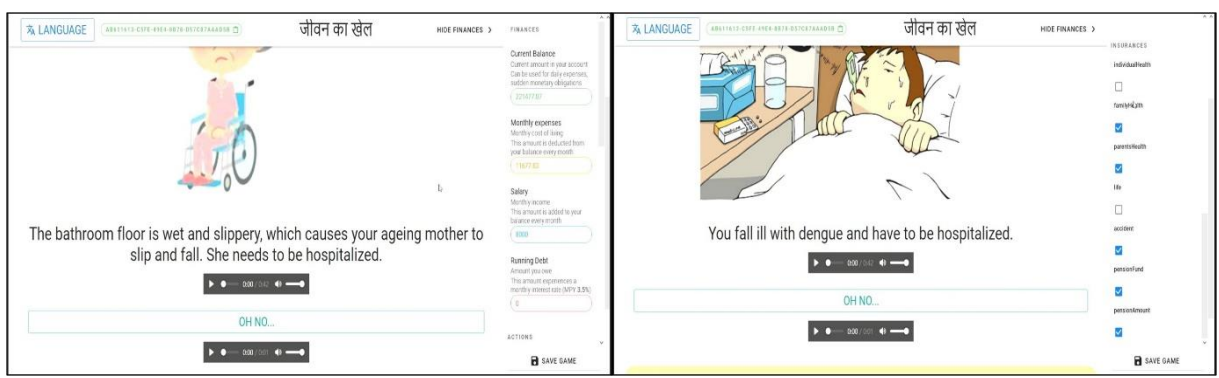

Figure 2d: The screenshot on the left presents an unfortunate accident to a family member requiring hospitalization and the one on the right is sickness effecting self. These are presented as 'loss-framing' conditions in the game for family health and individual health insurance products.

3.4 Hypotheses

In this study, we explore 2 main factors influencing willingness-to-purchase insurance products, the first is role of loss-framing. We test the following hypothesis:

*H1: loss-framing positively effects insurance purchase.*

Theoretical studies have shown that the demand for insurance increases with risk aversion ([76]; [77]) and risk-taking behaviour is an essential factor in insurance decision-making ([78]). Our premise is that individual with lower risk taking score or risk averse (from DOSPERT scores) translates to higher willingness to purchase insurance.

*H2: Insurance purchase has a positive correlation to risk averse attitude.*

4 | Results

4.1 Age, insurance and role of nudge

The participants are divided into two age groups (Group A: 18-25 years & Group B & C: 35-55 years). Though group C and group B are in the same age range, a separate label was considered as education and salary levels differed. As observed in (Figure 3), the younger age group is comparatively more inclined to purchase nearly all insurance products and the pension plan. Insurance and pension plans are more acceptable in group B compared to group C. It is interesting to note this difference between groups B & C, where the effect due to salary level, education and job prospects is evident across the insurance types and in pension plans. The lower percentage of insurance purchases in older adults discounts the tax deductions one can avail for insurance premium payments as per the country's taxation rules. Another probable explanation could be that the tax benefit gap is filled by other factors like home loan EMIs, tax brackets, or investments in the stock market that provide better long-term returns.

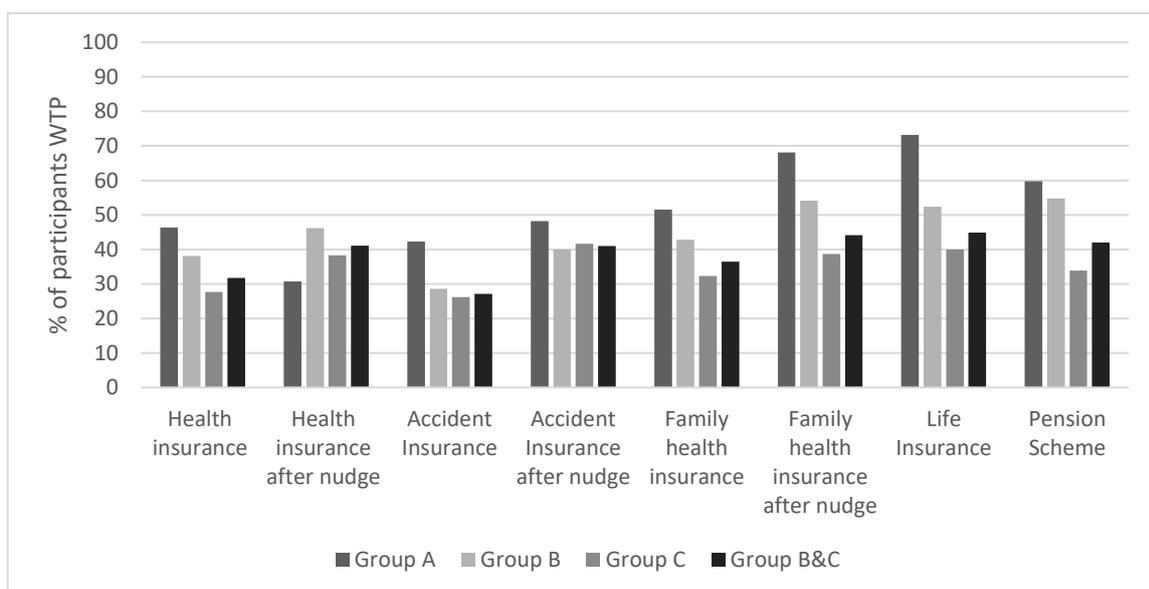

Figure 3: Shows descriptive statistics of the willingness-to -purchase percentage (y-axis) across the participant groups as a function insurance options.

The loss framing effect is evident in all demographics, except for a reversal trend in the younger population for individual health insurance. For family health insurance, the difference between before and after loss framing is a 19% rise for younger participants and a 4% increase for older participants; for accident insurance, the loss framing impact is 6% for younger and 22% for older individuals. The data supports our hypothesis (H1), but also demonstrates that the influence of loss framing is dependent on the insurance product. A 2-proportion test (Table 1) was applied to assess the influence of loss framing on each insurance product (within-subject) and between the two age groups. For this analysis, the number of participants who had insurance before the mishap (sample event shown in Fig.4d) was subtracted from the total number of participants for the estimate. The post mishap effect was similar for the accident and individual health insurance options in both age groups. All other insurance options show a significant difference (at $\alpha = 0.05$) as a function of age.

Table 1: The P-value score of the Two-Proportion test for the two age groups and the insurance options.

| Insurance option | Group A Proportion | Group B/C Proportion | P-Value |
|---|---|---|---|
| Individual Health Insurance | 0.46 | 0.33 | **0.028** |
| Individual Health insurance post mishap | 0.3 | 0.38 | 0.17 |
| Family Health insurance | 0.52 | 0.38 | **0.02** |
| Family Health insurance post mishap | 0.71 | 0.42 | **0.001** |
| Accident Insurance | 0.42 | 0.28 | **0.01** |
| Accident insurance post mishap | 0.48 | 0.5 | 0.4 |
| Life insurance | 0.72 | 0.47 | **0.0001** |
| Pension Scheme | 0.59 | 0.46 | **0.03** |

4.2 Risk Attitudes and insurance purchases

The responses to the DOSPERT questions (Financial – 6 questions and health – 3 questions) were analyzed for risk attitude metrics. Each question was rated on a one-to-seven Likert scale, with seven being 'extremely likely'. The Cronbach alpha to test for internal consistency across all questions was 0.75 (acceptable) for Group A, 0.86 for group B (good) and 0.85 for Group C. From the between group difference analysis (average score plot – Figure 4 left), statistical significance was present for the financial scale between group A & B (Mann Whitney U test: $z$-score is 2.14207, $p$-value = 0.03236 at $p < .05$). But the distribution in the scores across participants was significant (Figure 4 right Box-Whisker plot). While Group C has near equal distribution to the median value, group A & B finance scale show skew to the bottom quartile. Overall, the younger age group (A) show higher-risk attitudes with middle-aged and lower income levels (group B) being more risk averse. Though the social risk scale questions administered were only 3, it is interesting to observe the lower-risk scores in the younger group.

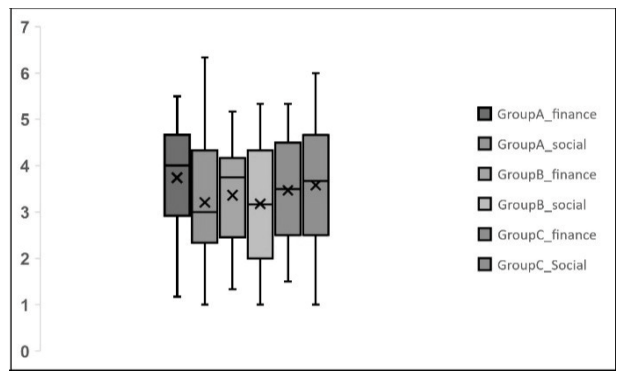

Figure 4: - The distribution for the financial and social risk scale of the DOSPERT survey

Logistic regression was applied to examine associations between risk-taking scores and willingness to purchase insurance for life insurance and pension plan. The average value of the risk survey response of each participant was taken as the variable. The participants in Group B/C self-disclosed loan responsibility in real life, which was considered in the analysis. The student-age population does not have loans in their names (parents or guardians usually pay the EMI); this data was not collected from Group A.

Panel data analysis was applied to investigate the correlation between taking risk-taking scores and willingness to purchase insurance, loans (groups B and C), and wealth level for time series data such as health insurance, accident insurance, and family health insurance, which are provided to participants twice before and after a loss.

Table 2: Shows the findings of a logistic regression used to determine a correlation between insurance options and DOSPERT risk scores for groups A & Group B/C (second and third column) and with loan for groups B and C combined (last column).

| Insurance | Correlation coefficients group A (P-value) | Correlation coefficients group B and C (P-value) | Correlation Coefficients Group B and C(P-value) with Loan |
|---|---|---|---|
| Life insurance | 0.326(0.161) | 0.215(0.473) | 0.517(0.416) |
| Pension Scheme | 0.048(0.819) | 0.224(0.472) | 1.109(0.090) |

Table 3: Shows the finding of panel data analysis to determine a correlation between insurance option, DOSPERT risk scores, and wealth level for group A for insurance option, which were asked multiple times to the participants. All significant values (p<0.05) are indicated in bold.

| Insurance | Correlation Coefficients Group A(P-value) with wealth | Correlation coefficients group A (P-value) with risk score |
|---|---|---|
| Health Insurance | **0.3469(0.00)** | 0.0606 (0.716) |
| Family Health insurance | **0.1234(0.030)** | - 0.2827 **(**0.061**)** |
| Accident Insurance | **0.1206(0.041)** | 0.1763 (0.260) |

Table 4: Shows the finding of panel data analysis to determine a correlation between insurance option, DOSPERT risk scores, loan and wealth level for groups B and C for insurance options which were asked multiple times to the participants.

| Insurance | Correlation | Correlation coefficients group | Correlation Coefficients |
|---|---|---|---|
| | | | |

|  | Coefficients Group B and C(P-value) with wealth | B and C (P-value) with Risk score | Group B and C(P-value) with Loan |
| --- | --- | --- | --- |
| Health Insurance | - 0.0781(0.273) | **- 0.3305 (0.022)** | 0.0185 (0.952) |
| Family Health insurance | 0.0742(0.187) | - 0.0644 (0.642) | 0.2872 (0.332) |
| Accident Insurance | - 0.0234(0.694) | **0.2915 (0.046)** | 0.3850(0.212) |

Table 3 shows that health insurance, accident insurance, and family health insurance positively correlate with Group A's desire to buy insurance and wealth level. For all insurance alternatives, there is no statistically significant correlation between risk scores and willingness to acquire insurance for group A members.

According to Table 4, only accident insurance has a statistically significant ($p < 0.05$) positive correlation with risk scores for Group B and C (older participants), whereas health insurance has a statistically significant negative correlation with risk scores. The positive correlation shows that as risk-taking grows, so does insurance consumption, whereas the negative correlation shows that as risk-taking increases, insurance consumption decreases. This renders hypothesis H2 for accident insurance invalid while validating hypothesis H2 for health insurance. For all insurance alternatives for groups B and C, there is no statistically significant correlation between having a financial responsibility in the form of a loan and desire to purchase insurance, as well as wealth level and willingness to purchase insurance.

## 5 | Discussion

We studied participants' willingness to purchase insurance using a novel paradigm. Setting the context mimicking real-life events in a game like interface with full information allowed for an experimental setup to study decision-making process. An immersive or engaging simulator with feedback is an advantage over survey-based responses. For ecological validity, the selected events and information (framing) presented were as experienced by a family. The data from a more comprehensive age range, especially the 30 – 50 year age group, is novel as most studies focus on 60+ age groups ([79];[80];[81]) or university students ([82]; [83];[84]). The 30-50-year cohort is critical to the insurance industry as it comprises populations in the workforce with varied education levels, salaries, family responsibilities, commitments (like loans and EMI's) and financial prospects. A comparative analysis of age, financial prospects, and behavioural aspects like individual risk attitude in insurance ([85]) have not been extensively studied; hence the findings from our study are a significant contribution. The main observation was the differential responses as a function of the insurance product, that is, savings products like life (endowment or whole life policy) and pension fared better as they are perceived as investment choices than health and accident insurance products.

5.1 Age, personal risk attitudes and willingness to purchase

It was interesting to observe that WTP is a function of the type of insurance and the subjective perception of risk or threat weighted by age and financial status reflected from real-life. The younger population (group A) with no immediate family responsibilities are inclined to purchase insurance products. We can attribute the group's potential financial prospects (the majority of the participants in this group were computer science students) and hence prospective affordability. This observation should be tempered with a possibility of acceptance of insurance products due to minimal real-life experiences (younger population) and the inability to realize the implications of the choices. Of interest are the differences between age-matched groups B & C, where financial prospects, income, and education seem to play a dominant role in decision-making. The lower uptake in the two groups can be attributed to misperceptions, optimism about the future, and overconfidence due to financial stability ([86]).

The mean values of the risk scores from the self-report DOSPERT Survey were slightly higher for the younger group. Only family health insurance (group A), Health insurance and accident insurance (groups B &C) correlations with risk attitudes show statistical significance testing hypothesis (H2). Group A (university student cohort), with higher risk-taking scores, show a higher WTP for all insurance categories supporting the findings ([87]) This contrasts with the hypothesis that younger age groups have a sense of longevity and hence do not consider accidents or health insurance ([88]). The result cannot be explained by classical expected utility theory, but as suggested by Eling et al. [89], it could also be due to not fully understanding the risk transfer mechanism. Group B, with higher risk aversion scores compared to the age-matched cohort (group C) show higher WTP, implying that lower salary, education levels, and financial prospects lead to investment in savings (whole life and pension) and health/accident insurance. The finding contrasts studies ([90]; [91]), which proposed that WTP is directly proportional to education, financial literacy, income or risk-aversion ([92]). Group B also shows an inclination for pension plans which can be attributed to financial protection by incremental investment to cover their post-retirement and ensure family protection. Group C, with a better economic position and education (compared to Group B), show the least WTP for all insurance products and pension plan. Furthermore, for groups B and C members who are tax-paying citizens, life insurance is frequently viewed as a risk-free form of tax-deferred investing ([93]).

Influence of loss aversion on insurance demand might also interact with narrow framing and subjective probability ([94];[95]) for group B. People with narrow frames are those whose preferences consider both consumption smoothing and gain-loss utility from prospect theory ([96]). While consumption smoothing increases insurance demand for risk-averse individuals, a concavity of gain-loss utility function leads to a negative correlation between risk aversion and insurance consumption, which could explain why risk aversion is negatively correlated to family health insurance. The non-student participant set also mentioned having personal loans in real life, though, as per our analysis, it does not seem to affect insurance choices significantly. While loans and the EMI one pays in real life determine purchasing insurance plans, the participants may have considered only the information provided in the application.

Analysing each insurance product, we notice that in the older population group (B &C), a positive and significant correlation value with risk attitude is observed for accident insurance and a negative significant correlation for health insurance. The findings support studies ([97]) that reported an increase in responsibility leads to an increase in risk aversion. A simpler explanation can be that the older groups were probable owners of vehicles, had been involved in accidents, and had made payments. The differential as a function of age could also mean probability distortions ([98];[99]) – wherein the older population underestimate health and over-estimate accidents.

Correlation analysis for life-insurance and pension plans with risk attitudes show no significant associations as also reported by Liebenberg et al. [100] . Support can also be found in the work of ([101];[102];[103]), who claim that savings may be a more important defining factor in insurance consumption than risk attitudes, and the return on life insurance is equally appealing to people with a varying willingness to take the risk. The study by Charness et al.[104] found that risk attitudes tested in the lab are not related to risk-taking in the field, calling into question the application of the same to understand decision-making under risk and uncertainty. It is possible that the domain-specific risk-taking scale poses hypothetical questions rarely encountered by the participants recruited for this study, hence a cultural and contextual factor weighing in. The differential behavior observed in our study contrasts previously reported works and can be attributed to the modality used for presenting the choice architecture, the inclusion of all insurance products presented with a context similar to real-life events.

5.3 Framing effect

Overall, the loss framing effect is evident for the three insurance products, validates our hypothesis (H1) and supported by theories on loss framing ([105];[106]). In the younger group, the loss framing is not evident only for individual health insurance (figure 3a) and as they are generally healthy and rarely experience health scares ([107]), the effect of loss framing effect seems limited. Older participants could have health problems, and faced financial loss, so the effect of loss framing is greater. In the case of accident insurance, both age groups show an increase in insurance consumption, with the older participants showing a greater rise because the overall cost due to accidents includes loss of employment or pay cuts in older participants and they are more likely to be severely injured ([108]). In the instance of family health insurance, the 2-proportion test reveals that the loss framing effect was comparable across age categories, owing to the similar impact of an unfortunate event across all groups.

In summary, as per expected utility theory ([109]; [110]), individuals will purchase insurance products as they understand the loss they would incur and usually risk-averse to changes in financial situation. Some studies demonstrate that risk increases insurance consumption ( [111];[112]), while others show the contrary where ambiguity is a factor ([113]). However, as can be inferred from the findings in our study, the decisions are not unconditional, with product offering playing a significant role. When the information/situation is presented as a form of sure loss, the participants are risk averse, which is integral to prospect theory ( [114];[115]).

**6 Conclusions**

We used a game-like application to investigate the impact of loss framing and personal risk attitude on WTP insurance. The focus was on two groups of participants: first, university students aged 18 to 25 with no personal source of income, and second, working-class people aged 35 to 55, a similar comparative analysis not presented in any of the previous studies in the country. The age-wise differential choices on insurance types are a major contribution, as laboratory experiments on insurance purchase have been on university students and non-academic analysis is focused on ages 50 and above. The working adult age with responsibilities is an important segment to understand as potential customers. Most studies infer risk aversion from the choice or willingness to purchase insurance product, while we also consider the inherent attitude towards risk-taking as a factor. The findings add to the existing literature on behavioral impact on risk-assessment, decision to purchase and role of constraints.

**7. Limitation**

In the absence of a systematic overview that shows the potential validity of game-based methods to simulate real-life decision making, it is prudent to consider limitations from gamification as applied in this experiment. Games are frequently associated with high cognitive load, and data collected from game-like experiments may limit the conclusions we can draw from the results ([116]). Typically, entertainment games give players a lot of say over what goals they pursue and what strategies they use Vorderer & Ritterfeld [117] and frequently results in disparate actions and outcomes from various players ([118]). Players generally expect complete control over which games they play and how long ([119]). However, in experimental settings where gamification is used to collect data, participants are typically assigned to predetermined conditions of gameplay, which may result in participants becoming frustrated because they force them to play games they dislike and would not normally play ([120]). Even though the gamification of data collection methods can only simulate certain aspects of reality ([121]; [122]), the results of such experiments can be applied to real-world problems ([123]). A second limitation is the sample size and gender consideration in each group. Third, information about current insurance policy consumption in real life was not considered. Fourth, the use of a general-purpose questionnaire (DOSPERT) for risk attitude measure as an index and does not consider situational risk and one's responses to it.